\documentclass[aps,prd,twocolumn,groupedaddress,showpacs,nofootinbib]{revtex4}
\usepackage{epsf,epsfig,graphics,graphicx}
\begin{document}

\title{Cosmological constant from gauge fields on extra dimensions}

\author{Hing-Tong Cho$^{1,2}$, Choon-Lin Ho$^1$, and Kin-Wang Ng$^2$}
\affiliation{$^1$Department of Physics, Tamkang University,
Tamsui, Taipei County, Taiwan 251, R.O.C.\\
$^2$Institute of Physics, Academia Sinica, Nankang, Taipei, Taiwan 115, R.O.C.}


\begin{abstract}

We present a new model of dark energy which could explain the
observed accelerated expansion of our Universe. We show that a
five-dimensional Einstein-Yang-Mills theory defined in a flat
Friedmann-Robertson-Walker universe compactified on a circle
possesses degenerate vacua in four dimensions. The present
Universe could be trapped in one of these degenerate vacua. With
the natural requirement that the size of the extra dimension could
be of the GUT scale or smaller, the energy density difference
between the degenerate vacua and the true ground state can
provide us with just the right amount of dark energy to account
for the observed expansion rate of our Universe.

\end{abstract}

\maketitle

It is now generally agreed that the biggest unsolved problem
in astronomy and cosmology is the newly discovered fact that our
Universe is undergoing a stage of accelerated expansion. Recent
astrophysical and cosmological observations such as structure
formation, type Ia supernovae, gravitational lensing, and cosmic
microwave background anisotropies have concordantly predicted a
spatially flat universe containing a mixture of matter and a
dominant smooth component, dubbed the ``dark energy'', which
provides an anti-gravity force to accelerate the cosmic
expansion~\cite{texas04}. The simplest candidate for this
invisible component carrying a sufficiently large negative
pressure for the anti-gravity is a true cosmological constant.
Current data, however, are consistent with a somewhat broader
diversity of such dark energy as long as its equation of state
approaches that of the cosmological constant at a recent epoch.
Many models have been proposed to account for this dynamical dark
energy, including quintessential and k-essential models, modified
gravity and scalar-tensor theories, and so on~\cite{dem}. In these
models, dark energy is characterized by an equation of state
$w(z)$ which is the ratio of its pressure and energy density at
redshift $z$. The cosmological constant is indeed vacuum energy
with $w(z)=-1$. Current observational data, when fitted to dark
energy models with a static $w$, favor the so-called $\Lambda$CDM
model with vacuum energy $\Omega_\Lambda=0.7$ and cold dark matter
$\Omega_{\rm CDM}=0.3$, while constraining
$w=-1.02^{+0.13}_{-0.19}$ at the $95\%$ confidence level.
Furthermore, joint constraints on both $w(z)$ and its time
evolution $dw(z)/dz$ at $z=0$ are consistent with the value of the
equation of state expected of a static cosmological
constant~\cite{riess}. However, the data are not precise enough to
pinpoint whether the dark energy is truly static or dynamical.
Most likely, new kinds of measurements or next-generation
experiments are needed to reveal the nature of dark energy.
In this paper, we will address the dark energy problem by
considering non-Abelian gauge theories on compact extra
dimensions.

Interest in theories of extra dimensions has been quite immense in
recent years~\cite{csaki}.  New models such as brane scenarios,
large extra dimensions, and warped extra dimensions have not only
revolutionized the Kaluza-Klein theory, but also shed new light on
some long-standing problems in particle physics and cosmology.
Interestingly, theories with large extra dimensions can be even
tested by future collider experiments. Recently cosmological
models involving extra dimensions have been constructed to account
for the current cosmic acceleration or to accommodate the dark
energy~\cite{models}. Here we shall show how a
five-dimensional (5d) gauge theory based on the Hosotani symmetry
breaking mechanism ~\cite{hoso} could naturally give rise to a
small but finite cosmological constant that plays the role of dark
energy.

The Hosotani mechanism is a non-Higgs-type symmetry-breaking
mechanism which has been widely discussed in the
literature~\cite{hoso2}. Its main idea is that in a multiply
connected spacetime manifold, the vanishing of the field strength
$F_{MN}=0$ of gauge fields $A_M$ in a vacuum does not necessarily
imply the vanishing of the gauge fields, and $A_M\neq 0$ will
imply gauge symmetry breaking in general.
The relevant order
parameter is the path-ordered Wilson lines
$U_n=P\exp\left(\oint_{C_n}A_Mdx^M\right)$, where $C_n$ represent
non-contractible loops in the manifold. If $U_n$ do not belong to
the center of the group, then the original gauge group is broken
to some subgroup that commutes with all $U_n$. This mechanism has
been extensively employed in superstring phenomenology~\cite{can},
and applied to Kaluza-Klein cosmology in connection with the
problem of vacuum stability~\cite{ho}. Recently, a new
extranatural inflation model in which the inflaton is the fifth
component of a gauge field in a 5d theory compactified on a circle
was presented~\cite{nima}, and it was shown that the fifth
component may also be a good candidate for quintessence if the
quintessential potential is provided by massive bulk fields with
bare masses of order of the GUT scale~\cite{pilo}.

In Ref.~\cite{ho}, the authors considered a 5d Einstein-Yang-Mills
theory, with massless fermions, defined in a flat
Friedmann-Robertson-Walker universe compactified on a circle. The
spacetime metric is given by
\begin{eqnarray}
ds^2&=&g_{MN}dx^M dx^N \quad\quad (M=0,1,2,3,5) \nonumber\\
    &=&dt^2-a^2(t)d{\vec x}^2-b^2(t)dx_5^2,
\label{metric}
\end{eqnarray}
and the action is consisted of gravity, $SU(2)$ gauge fields, and
a fermionic sector $L_f$ which contains $N_f$ massless gauged fermions and
$n_f$ massless free fermions:
\begin{equation}
S=-\int d^{5}x {\sqrt{|g_{MN}|}}\left(\frac{R}{16\pi{\bar G}}
+{1\over2}{\rm Tr} F_{MN}F^{MN} + L_f \right).
\label{action}
\end{equation}
Here the gauge bosons and the fermions can be considered as fields 
in some hidden sector of a grand unified theory.
Dynamics of the gauge fields and the fermion fields here
determines the stucture of the vacuum.
The Casimir energy of the system was computed by evaluating
the one-loop effective potentials of the gauge fields and the
fermions in the backgrounds defined by the metric~(\ref{metric})
and the classical gauge fields of the form
\begin{equation}
A_\mu=0,\quad A_5=\phi(t){\bf\sigma}^3,\quad \mu=0,1,2,3\,,
\end{equation}
where $A_5$ is the fifth component along the circle and
${\bf\sigma}$ is the Pauli matrix. This amounts to a total
effective potential of the system denoted by $V(b R_0,\phi)$,
where $R_0$ is the final radius of the circle. It was
shown that the Einstein's equations for $a(t)$, $b(t)$
and $\phi(t)$, derived from the action~(\ref{action}), admit
static vacuum solutions with stable compactification. These
solutions with $\dot a_0=\dot b_0=\dot\phi_0=0$ are given by the
global minima of $V(b R_0,\phi)$ determined by
\begin{equation}
2g_5 b_0 R_0\phi_0=\frac{r}{2}~~~{\rm (mod}~r), \label{vacuum}
\end{equation}
where $g_5$ is the 5d $SU(2)$ gauge coupling and $r=1$ and $r=2$
correspond respectively to periodic fermions in the adjoint and
fundamental representations. In the case of adjoint fermions, the
vacuum states correspond to $U(1)$ symmetry, hence $SU(2)$ gauge
symmetry is dynamically broken. 
For fundamental fermions the
gauge symmetry is unbroken as the Wilson line is an element of
$Z_2$. To obtain a zero cosmological constant for these vacuum
states, an appropriate number ($n_f$) of free fermions with
assigned boundary conditions along the circle has been chosen such
that $V(b_0 R_0,\phi_0)=0$. A salient feature of this model is
that only a small number of fermions is required to stabilize the
vacuum. In fact, $N_f=1$ gauged fermion will suffice to do the
job. On the contrary, in the other Kaluza-Klein-type theories one
usually needs to add a large number (of the order of $10^4$) of
matter fields for the same purpose.

Now we turn to the dark energy problem. Naively, one would expect
that the vacuum energy density $\rho_\Lambda$ is of order $M_P^4$,
where the reduced Planck mass is given by $M_P=(8\pi
G)^{-1/2}=2.44\times 10^{18} {\rm GeV}$, since the Planck scale is
the natural cutoff scale of zero-point energies of each quantum
field. But the observed value for the vacuum-like energy density
is $\rho_\Lambda \simeq 0.7\rho_c \simeq 1.6h^2\times10^{-120}
M_P^4$, where $h\simeq 0.7$ is the present Hubble parameter
defined by $H_0=8.76h\times 10^{-61} M_P$~\cite{texas04}, and so
the naive estimate is larger than the observed value by a factor
of $10^{120}$. Many solutions have been proposed about a
vanishingly small cosmological constant in some ultimate ground
state~\cite{vil}. Here we assume that the cosmological constant
absolutely vanishes in a true ground state with lowest possible
energy density. Then, we will show that the vacuum states in the
model considered above are indeed metastable due to quantum
tunneling effects and will eventually settle down to this true
ground state. If the present Universe is still in one of these
quasiground states, then the energy density difference above the
true ground state can provide us with a small but finite
cosmological constant. This idea has already been put forth by
considering the topological vacua in a 4d $SU(2)$ Yang-Mills-Higgs
theory which is spontaneously broken to $U(1)$ via the Higgs
mechanism with a Higgs potential~\cite{yoko}. Unfortunately, the
existence of massless or light gauged fermions would suppress the
tunneling~\cite{call} and thus spoil the idea. Although the
fermions can get masses via the Higgs mechanism, the Yukawa
couplings are quite arbitrary and may be very small.  In contrast,
our scenario has new merits. Firstly, we do not need to introduce
an {\it ad hoc} Higgs field, which is here replaced by the fifth
component of the gauge field. Secondly, the gauge symmetry
breaking needs not be introduced arbitrarily, but instead is
determined dynamically by the Casimir energy of the non-integrable
phase of the gauge field on the compact dimension through the
Hosotani mechanism.  Thirdly, the gauged fermions naturally get
huge masses of order of the inverse of the size of the compact
extra dimension.  Lastly, the cosmological constant in our model,
as we will show below, is naturally related to the size of the
compact extra dimension.

Without loss of generality, let us consider a universe
corresponding to the vacuum state in the case of periodic
fundamental fermions with $2g_5 R_0 \phi_0=1$, where
we have set $a_0=b_0=1$ and $r=2$ in Eq.~(\ref{vacuum}).
Presumably, the Universe has undergone the compactification
of the circle with initial conditions at some early time $t_i$
(for instance, given by $a(t_i)$, $b(t_i)$, and $\phi(t_i)\simeq 0$),
and rolls down the effective potential $V$ to this vacuum state.
The actual evolution is very interesting and it warrants a detailed
study of the Einstein's equations. In fact, this scenario has been
discussed in the context of the extranatural inflationary model~\cite{nima}.
Here we only concern about the vacuum state and the dark energy problem.
At energies below $1/R_0$, the 4d
effective action for the zero Fourier modes of gauge fields in
Eq.~(\ref{action}) is given by
\begin{equation}
S^{\rm gauge}_{\rm eff}=-\int d^4 x \left({1\over2}{\rm Tr}{\tilde
F_{\mu\nu}}{\tilde F^{\mu\nu}}+ {\rm Tr}{\tilde F_{\mu 5}}{\tilde
F^{\mu 5}} \right).
\end{equation}
Note that we have rescaled $A_M={\tilde A_M}/{\sqrt{2\pi R_0}}$
and $g_5=g_4{\sqrt{2\pi R_0}}$, where $g_4$ is the dimensionless
4d $SU(2)$ gauge coupling. Since ${\tilde
A_\mu}$ is independent of $x_5$ in the effective action, so
$\partial_5 {\tilde A_\mu}=0$ and ${\tilde F_{\mu5}}$ reduces to
the covariant derivative of ${\tilde A_5}$. Hence, by
rewriting ${\tilde A_5}=\Phi$, we have
\begin{equation}
S^{\rm gauge}_{\rm eff}=\int d^4 x \left(-{1\over2}{\rm Tr}{\tilde
F_{\mu\nu}} {\tilde F^{\mu\nu}} +{\rm
Tr}(D_\mu\Phi)(D^\mu\Phi)\right). \label{higgs}
\end{equation}
This resembles the $SU(2)$ Yang-Mills-Higgs system without a Higgs
potential. 
Although the 5d gauge symmetry is unbroken for the vacuum
states given by Eq.~(\ref{vacuum}), the fifth component obtains a
non-zero vacuum expectation value
$\langle\Phi\rangle={\tilde\phi_0}{\bf\sigma}^3$, where
${\tilde\phi_0}=1/(2g_4 R_0)$~\cite{ho}. 
Thus, the zero-mode gauge fields and gauged fermions
acquire huge mass equal to $1/(2R_0)$~\cite{mass}. 
In the case of
periodic adjoint fermions, the system is spontaneously broken to
$U(1)$ by the Wilson loop along the small circle with
$\langle\Phi\rangle={\tilde\phi_0}{\bf\sigma}^3$, where
${\tilde\phi_0}=1/(4g_4 R_0)$~\cite{ho}. Nevertheless, some
fermion modes still remain massless~\cite{mass}. Since these
massless fermions suppress the vacuum
tunneling~\cite{call}, we will not consider this case.

It is well known~\cite{call,thooft} that a non-Abelian gauge
theory, such as that described by Eq.~(\ref{higgs}), bears
degenerate perturbative vacua classified in terms of the winding
number $n$ and denoted by $|n\rangle$. Before we go on, we should
emphasize that there are no instantons nor topological numbers for
the 5d gauge fields in Eq.~(\ref{action}). Even in the massless
sector (given by Eq.~(\ref{higgs})), we still do not have exact
instantons but only ``constrained ones". However, these
constrained instantons are sufficient in describing the tunneling
phenomenon~\cite{thooft,aff} as we shall do in due course. Thus
let us consider the 4d $SU(2)$ Yang-Mills theory with gauge
coupling $g$. While the degenerate vacua are each separated by an
energy barrier, the change of the winding number can take place
through quantum tunneling from one vacuum to another. As such, the
true ground state, so-called the $\theta$ vacuum, can be
constructed as
\begin{equation}
|\theta\rangle=\sum_{n=-\infty}^{\infty}e^{in\theta}|n\rangle,
\end{equation}
where $\theta$ is a real parameter.
Using the dilute instanton gas approximation~\cite{call},
it was found that the expectation value of the Hamiltonian $\cal
H$ over the Euclidean spacetime volume $VT$ is given by
\begin{eqnarray}
\langle\theta|e^{-{\cal H}T}|\theta\rangle &\propto&
\sum_{n,n'}\langle n'|e^{-{\cal H}T}|n\rangle e^{i(n-n')\theta}
\nonumber \\ &=&\sum_
{n,n'}\frac{1}{n!}\frac{1}{n'!}(KVTe^{-S_0})^{n+n'}
e^{i(n-n')\theta} \nonumber \\ &=&\exp(2KVTe^{-S_0}{\rm
cos}\theta),
\label{tvac}
\end{eqnarray}
where $S_0=8\pi^2/g^2$ is the Euclidean action for an instanton
solution which corresponds to the quantum tunneling from
$|n\rangle$ to $|n\pm 1\rangle$. Here $K$ is a positive
determinantal prefactor. Eq.~(\ref{tvac}) shows that the energy density
difference between each $\theta$ vacuum and the perturbative vacuum is
given by
\begin{equation}
\langle\theta|\rho_\Lambda|\theta\rangle
- \langle n|\rho_\Lambda|n\rangle = -2Ke^{-S_0}{\rm cos}\theta.
\end{equation}
Therefore, the $\theta$ vacuum with the lowest energy is given by $\theta=0$.
Let us assume that the $\theta=0$ vacuum is the true ground state where
the vacuum energy vanishes. In fact, it has been proposed~\cite{nielsen}
that the effect of wormholes~\cite{cole} drives the Universe to this
CP-symmetric $\theta=0$ vacuum state with zero cosmological constant.
(More discussions about this can be found in Ref.~\cite{yoko}
and references therein.) Then, after normalizing
$\langle\theta=0|\rho_\Lambda|\theta=0\rangle=0$, we find that the
vacuum energy density in each perturbative vacuum is
\begin{equation}
\langle n|\rho_\Lambda|n\rangle= 2Ke^{-S_0}.
\end{equation}
This vacuum energy will manifest as a cosmological constant
provided that we still live in one of these perturbative vacua. To
guarantee this condition, the tunneling probability from any one
of these perturbative vacua to the true ground state in the
current horizon volume in the cosmic age must satisfy $\Gamma
H_0^{-4}\le 1$, where $\Gamma$ is the tunneling rate per unit
volume per unit time given by $\Gamma\simeq \rho_\Lambda
e^{-S_0}$.\footnote{It has been proved~\cite{cole2} that the
$|n\rangle$ vacua are not physical because they do not satisfy
cluster decomposition: $\langle n|e^{-{\cal
H}(T_1+T_2)}|n+\nu\rangle=\sum_p \langle n|e^{-{\cal
H}T_1}|p\rangle \langle p|e^{-{\cal H}T_2}|n+\nu\rangle$ for large
times $T_1$ and $T_2$. Here, however, the tunneling time is longer
than the age of the Universe. So, observationally we have $\langle
n|e^{-{\cal H}T}|p\rangle \simeq \delta_{np}$, which implies the
cluster decomposition, $\langle n|e^{-{\cal
H}(T_1+T_2)}|n\rangle=\langle n|e^{-{\cal H}T_1}|n\rangle \langle
n|e^{-{\cal H}T_2}|n\rangle$.}

Unfortunately, for a pure $SU(2)$ gauge theory which is
scale-invariant, the size of the instanton $\rho$ is arbitrary,
and so the prefactor $K$ which involves an integral over instanton
sizes diverges as $\rho$ goes to infinity. However, if the $SU(2)$
gauge field is coupled to a Higgs scalar with isospin $q$ and
vacuum expectation value $v$, the quantum tunneling will proceed
via a constrained instanton whose size is cut off at a scale
$v^{-1}$~\cite{thooft,aff}. The Higgs contribution to the Euclidean
action of the constrained instanton is approximately given
by $S_H=4\pi^2q\rho^2 v^2$, rendering $K$ a finite quantity.
Including $N_f$ gauged fermions with mass $m_f$, we find that~\cite{thooft}
\begin{eqnarray}
K&=&\frac{4\pi^2}{\alpha^4} \int_0^\infty \frac{d\rho}{\rho^5}
(m_f\rho)^{N_f} \exp\left[-S_H+c_1\ln(\rho v)+c_2\right] \nonumber \\
&=& 2\pi^2 e^{c_2}
\frac{\Gamma[(N_f+c_1-4)/2]}{(4\pi^2q)^{(N_f+c_1-4)/2}}
\left(m_f\over v\right)^{N_f} \left({v\over \alpha}\right)^4,
\label{kfactor}
\end{eqnarray}
where $\alpha=g^2/(4\pi)$ is the gauge coupling strength at the energy
scale $v$, $c_1=20/3-2N_f/3$, and $c_2=5.96-0.36N_f$.
Therefore (c.f.~\cite{yoko}), if $m_f=gv$, we will have
\begin{eqnarray}
\rho_\Lambda&\simeq& c^2\left({v\over \alpha}\right)^4
\alpha^{N_f\over 2} e^{-2\pi/\alpha}, \\
\Gamma &\simeq& c^2\left({v\over \alpha}\right)^4
\alpha^{N_f\over 2} e^{-4\pi/\alpha},
\end{eqnarray}
where
\begin{equation}
c^2=4\pi^2 e^{c_2} (4\pi)^{N_f\over 2}
\frac{\Gamma[(N_f+c_1-4)/2]}{(4\pi^2q)^{(N_f+c_1-4)/2}}.
\end{equation}
The requirements that $\rho_\Lambda=1.6h^2\times10^{-120} M_P^4$
and $\Gamma H_0^{-4}\le 1$ give
\begin{eqnarray}
{\pi\over{2\alpha}}+\left(1-{N_f\over 8}\right)\ln\alpha&=&
\ln\left(\frac{v}{M_P}\right)+30\ln10 \nonumber \\
&&-{1\over2}\ln\left(\frac{1.27h}{c}\right),
\label{rho}\\
\frac{v}{M_P}&\ge& 1.45\alpha^{1-{N_f\over 8}}/\sqrt{c}.\label{gamma}
\end{eqnarray}
For the minimal value of $v$ in Eq.~(\ref{gamma}) and $h=0.7$, we
obtain $\alpha\simeq 1/44.25$. Let us assume that $q=1$. Then, we
have $v\simeq 3.8\times 10^{16} {\rm GeV}$ for $N_f=1$ and
$v\simeq 8.3\times 10^{16} {\rm GeV}$ for $N_f=3$. If we take
$v=M_P$ in Eq.~(\ref{rho}), we will obtain $\alpha=1/46.9$ for
$N_f=1$ and $\alpha=1/46.4$ for $N_f=3$

To apply the above results to our model~(\ref{higgs}), we make the
replacements $g=g_4$ and $v={\tilde\phi_0}=1/(2g_4 R_0)$. Thus, we
find that $R_0^{-1} \ge 4.1\times 10^{16} {\rm GeV}$, which
indicates that the maximal size of the compact extra dimension is
of the order of the GUT scale. This result is obtained without
fine-tuning. The smallness of the cosmological constant is related
to the tunneling rate between the various vacua of the theory
which is also very tiny. This idea was first put forth generically
in Ref.~\cite{yoko} based on the Higgs mechanism, but
Eq.~(\ref{kfactor}) shows that the existence of massless or light
fermions with $m_f\ll v$ will suppress the prefactor $K$.
Consequently, the $|n\rangle$ vacua do not satisfy cluster
decomposition due to the chirality conservation~\cite{call}. Here
we revolutionize the idea by replacing the Higgs field with the
gauge field in the extra dimension through the Hosotani mechanism.
The vacuum state is chosen by the minimum of the Casimir energy
instead of the rather {\it ad hoc} Higgs potential. An important
advancement is that the gauged fermions in our model naturally
obtain a huge and definite mass given by $m_f=gv$. To our
knowledge, the Hosotani mechanism is the only way that does the
job.

In summary, we have constructed a new model of dark energy based on
a 5d Einstein-Yang-Mills theory. The presence of dark energy comes
out rather naturally in our model just from the assumption that
there exist extra dimensions. Symmetry breaking by the Hosotani
mechanism and the constrained instantons related to the vacuum
structure of the gauge field are both immediate consequences. As
long as the size of the extra dimension is between the GUT and the
Planck scales, the resulting cosmological constant will then be
just of the right amount to account for the dark energy content in
our Universe. No other ingredients are needed to achieve this goal.

This work was supported in part by the National Science Council,
Taiwan, ROC under the Grants NSC 94-2112-M-032-009 (H.T.C.), NSC
94-2112-M-032-007 (C.L.H.), and NSC 94-2112-M-001-024 (K.W.N.).

\end{document}